\documentclass{webofc}
\usepackage[varg]{txfonts}   % Web of Conferences font
\usepackage{hyperref}
\usepackage{url}
\hypersetup{colorlinks=true,citecolor=blue,urlcolor=blue,linkcolor=blue}
\usepackage{subfigure}

\usepackage{braket}

\newcommand{\modes}{J}
\newcommand{\modesq}{\modes_q}
\newcommand{\modesa}{\modes_{\overline{q}}}
\newcommand{\modesg}{\modes_g}

\newcommand{\occf}{{\omega}}
\newcommand{\occb}{{\widetilde{\omega}}}
\newcommand{\occmax}{\occb^{\mathrm{max}}}

\begin{document}

\title{Quantum Computation for Jets in Heavy Ion Collisions}

\author{
\firstname{Wenyang} \lastname{Qian}
\inst{1}\fnsep\thanks{\email{qian.wenyang@usc.es}}
}
\institute{
Instituto Galego de F{\'{i}}sica de Altas Enerx{\'{i}}as,  Universidade de Santiago de Compostela, Santiago de Compostela 15782, Galicia, Spain
}
\abstract{
Quantum computing has recently emerged as a transformative tool for investigating the real-time dynamics of jets in heavy-ion collisions, offering novel approaches to simulate non-equilibrium processes and strongly coupled phenomena that are challenging for classical methods. Here, I summarize my talk at Hard Probes 2024 at Nagasaki.
}
\maketitle

\section{Introduction}\label{sec:introduction}
% \wq{HIC state of art, jet, QC developments, more computing, why is it necessary, and three frameworks (1 page)}

Heavy-ion collisions (HIC) represent a frontier in experimental nuclear physics, wherein collisions of heavy nuclei at ultra-relativistic speeds create extreme conditions similar to those of the early universe. Facilities such as the Large Hadron Collider (LHC) and the Relativistic Heavy Ion Collider (RHIC) have provided an unprecedented window into the formation and evolution of the quark-gluon plasma (QGP). The state-of-the-art in HIC experiments involves sophisticated detectors and high-precision data acquisition that capture the complex dynamics of jet quenching, particle correlations, and phase transitions inherent in QCD matter.

Parallel to these experimental advances, quantum computing (QC) has emerged as a transformative technology poised to address some of the computational challenges inherent in simulating such complex quantum systems. Classical simulations of quantum field theories, including the dynamics of jets in a QGP medium, are hampered by the exponential growth of the Hilbert space with system size. By leveraging the principles of superposition and entanglement, QC is potentially advantageous in simulating many-body quantum systems. Recent developments in quantum information science and its associated hardware have spurred remarkable progress in quantum simulation in high-energy physics; For a review about the recent progress, see Refs.~\cite{Bauer:2022hpo, DiMeglio:2023nsa, Banuls:2019bmf}.

In this context, several QC frameworks are being actively explored. Digital quantum computing, that utilizes quantum circuits composed of unitary gates, allows for the implementation of algorithms such as the Variational Quantum Eigensolver (VQE) and Quantum Phase Estimation to approximate the spectra of complex Hamiltonians. Analog quantum computing, on the other hand, employs engineered physical systems—such as ultracold atoms or trapped ions—to mimic the dynamics of a target Hamiltonian directly. Furthermore, tensor network methods have been extensively used in classical simulations as a robust and promising framework for simulating quantum many-body systems, such as approximating ground states and low-entanglement regimes. These simulation approaches offer distinct advantages and limitations in terms of control, error rates, and scalability, and all aim towards inherently capture the exponential complexity of QCD phenomena.

% Recent advances in quantum hardware and algorithm design have opened up new avenues for simulating physical processes that are intrinsically quantum mechanical. As Richard Feynman famously stated, “Nature isn’t classical, dammit,” underscoring the need to employ quantum simulation for an accurate representation of natural processes~\cite{feynman1982simulating}. Here, we discuss how the rapid development in quantum computing, from noisy intermediate-scale quantum devices to prospective fault-tolerant machines, paves the way for novel approaches in both experimental and theoretical studies of heavy-ion collisions.

% Furthermore, the vast amounts of data produced in experiments like those at the Large Hadron Collider (LHC) necessitate innovative computing methods. The integration of quantum algorithms—such as Grover’s search and quantum annealing techniques—into the analysis pipeline promises significant speedups in data processing and pattern recognition, thereby offering fresh insights into jet clustering, tracking, and the fundamental dynamics governing heavy-ion collisions.

\section{Quantum computing in heavy-ion collisions}\label{sec:qc-hic}

QC techniques are well-motivated for heavy-ion collision experiments. QC can assist in processing experimental data, particularly in tasks such as particle tracking and jet clustering by transforming the task into Quadratic Unconstrained Binary Optimization (QUBO) problems~\cite{Pires:2020urc}, together with quantum search algorithms offering a theoretical speedup.
The integration of quantum algorithms to tackle jet clustering has shown promise: utilization of QUBO and Grover’s search strategies, for instance, allows for improvement on calculation of an event shape called thrusts~\cite{Delgado:2022snu}. The application of quantum machine learning may also extends the capabilities of traditional data analysis by encoding classical data into quantum states to potentially extract subtle features from high-dim datasets. This includes tasks like anomaly detection, where parameterized quantum circuits and autoencoders are employed, to flag unusual events in heavy-ion collisions. Novel applications are found for b-jet identification to simulated data of LHCb experiments~\cite{Gianelle:2022unu} with more underway.

On the theory side, there are vast number of application of QC to various stages of the HICs, ranging from hadron structures to scattering processes.
For example, QC is a new tool for determining hadron structure by directly extracting non-perturbative quantities such as parton distribution functions and hadronic tensors~\cite{Lamm:2019uyc}. While traditional lattice QCD methods struggle to compute these real-time correlators, quantum simulation integrating state preparation and light-front correlator evaluation are able to overcome this difficulty, with proof-of-concept studies in simplified (1+1)-d field theories on ideal quantum circuits and tensor networks~\cite{LiTianyin:2021kcs, Banuls:2024oxa, Kang:2025xpz}.
Further efforts are made in studying fragmentation functions~\cite{Grieninger:2024axp, LiTianyin:2024nod}.
Simulating real-time scattering processes is critical for understanding jet evolution, heavy-quark transport, and particle interactions within the QGP. Frameworks utilizing effective theories illustrate how QC can be used to simulate dynamics of the low energy from first principles~\cite{Bauer:2021gup}. Quantum walk may be used to simulate parton showers efficiently~\cite{Bepari:2021kwv}. Real-time dynamics of string breaking in QED is revisited and reinterpreted for high-energy phenomena~\cite{Hebenstreit:2013baa,Florio:2023dke}. Hadron states are prepared and evolved successfully on large scale quantum devices with 112 qubits~\cite{Farrell:2024fit}, which looks promising for future developments.

\section{Quantum simulation of jets in heavy-ion collisions}\label{sec:qc-jet}
% \wq{available methods, our approaches (method, strength), available works, future directions (3-4 pages), energy-energy correlator, look at Joao's new paper}
Experimentally, the properties of the QGPs in the ultra-relativistic high-energy collisions can be indirectly extracted by studying the yield and properties of QCD jets, i.e., beams of high-velocity particles generated during the same collisions, from cold nuclear matter to the hot quark-gluon plasma. Jets are complicated objects for their multi-particle nature and medium interactions. The success of the HIC program relies on having a clear handle on the jet evolution and its medium-induced modifications~\cite{Casalderrey-Solana:2007knd}.
Traditional interpretation of the jets’ modifications admits a classical description. In recent years, several quantum simulation strategies for studying jets or hard probes in general have been proposed and developed:

\textbf{Lattice QFT} continues to provides first-principle description of jet phenomena in the Hamiltonian formulation thanks to QC. The (1+1)-d QED (Schwinger model) are primarily used as a benchmark, along with (2+1)-d SU(2) models. Prominent applications are found in studying 
% vacuum modification due to propagating jets~\cite{Florio:2023dke}, 
jet energy loss~\cite{Barata:2025hgx}, energy-energy correlator~\cite{Barata:2024apg,Lee:2024jnt} and so forth. 

\textbf{Open quantum system} is another promising avenue of investigation for non-equilibrium dynamics. By describing the hard probes in the QGP through Lindblad equation, the probe's information can be extracted by tracing out the environment~\cite{DeJong:2020riy, deJong:2021wsd}. Various novel quantum algorithms are also proposed to boost the simulation efficiency.

\textbf{Light-front Hamiltonian formalism} stands out as a dominant approach to directly describe the jet's interaction with the medium. On the light-front, the full (3+1)-d QCD Hamiltonian is used to describe the dynamics of the probes, together with a classical background field for the medium. Various efforts are made over the years gearing towards more efficient simulation at large scale~\cite{Barata:2021yri,Barata:2022wim, Barata:2023clv, Qian:2024gph, Wu:2024adk, Yao:2022eqm,Castro:2025ocx}, which I will focus in my talk below. 

\subsection{Light-front Hamiltonian approach}
\begin{figure}
    \centering
    \subfigure[\label{fig:spacetime}]{\includegraphics[height=0.35\textwidth]{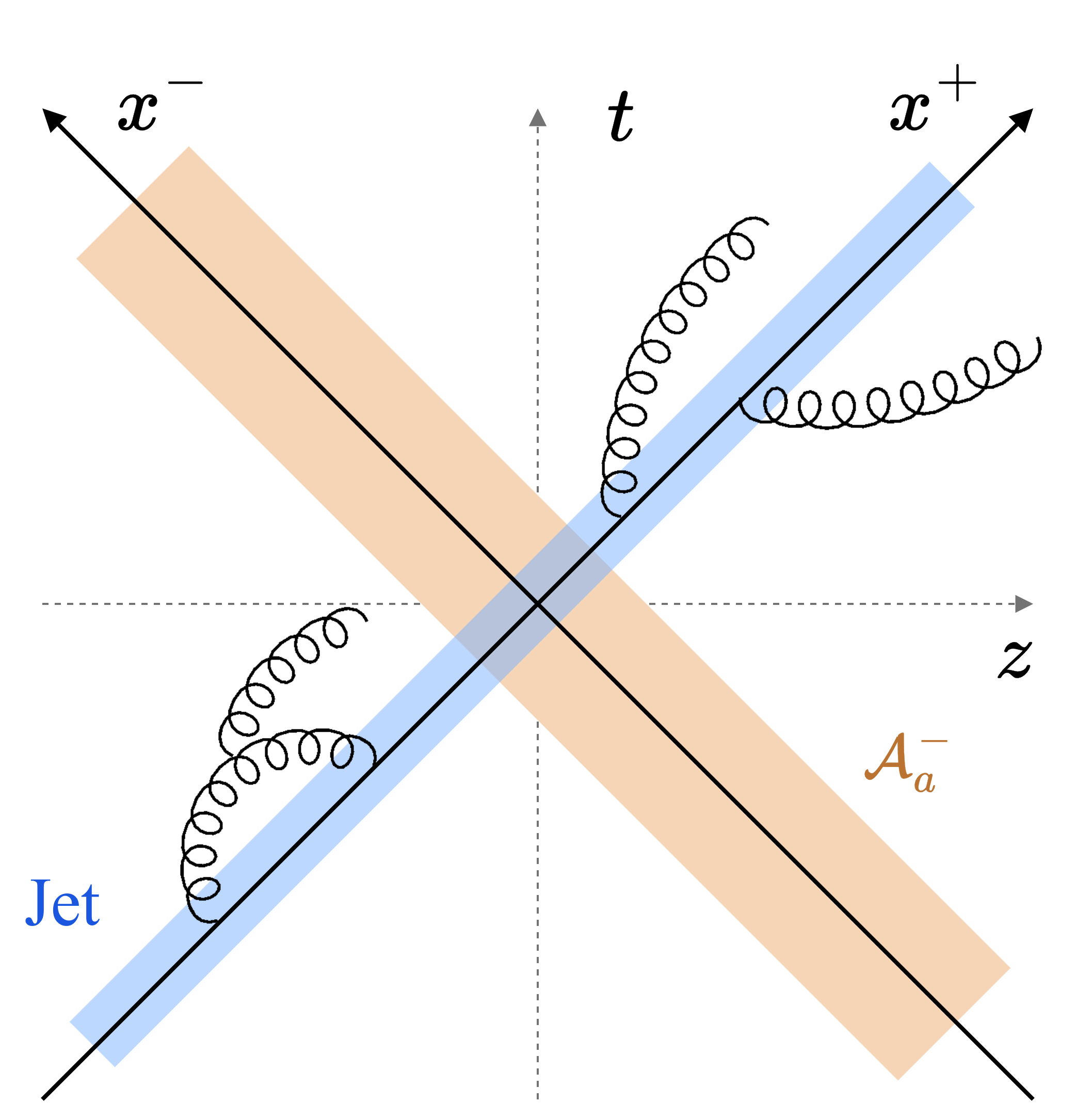}
    }\quad\quad
    \subfigure[\label{fig:interference}]{\includegraphics[height=0.30\textwidth]{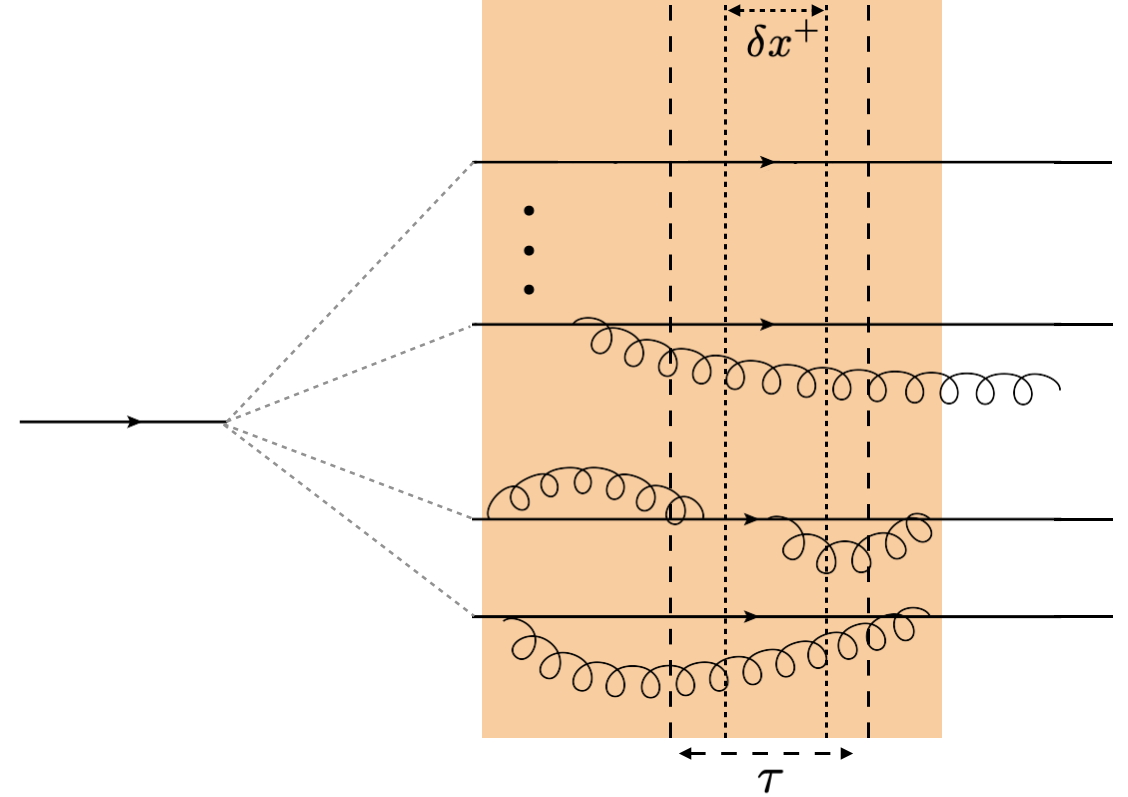}
    }\\
    \subfigure[\label{fig:encoding}]{\includegraphics[width=0.45\textwidth]{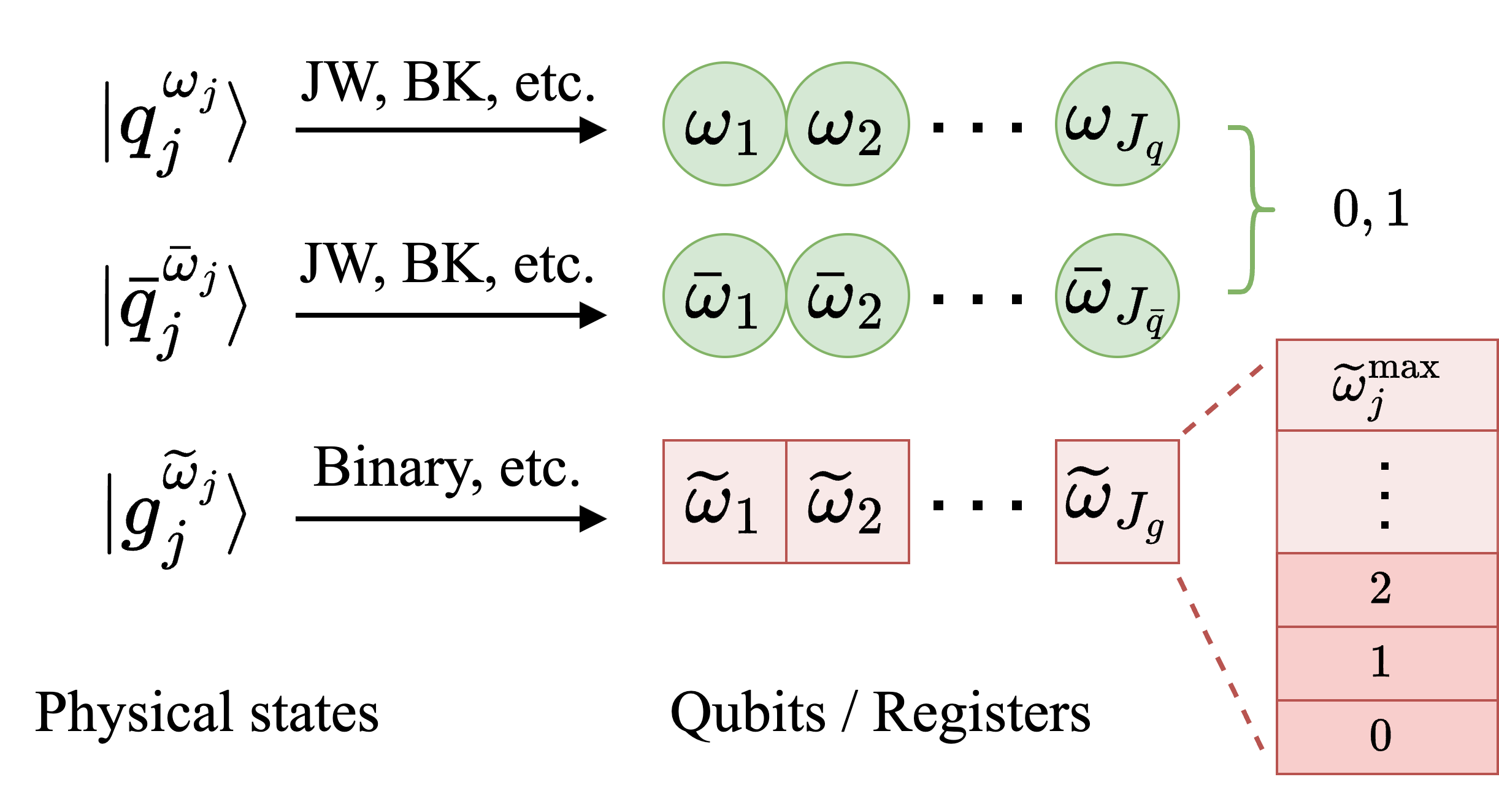}
    }\quad
    \subfigure[\label{fig:circuit}]{\includegraphics[width=0.48\textwidth]{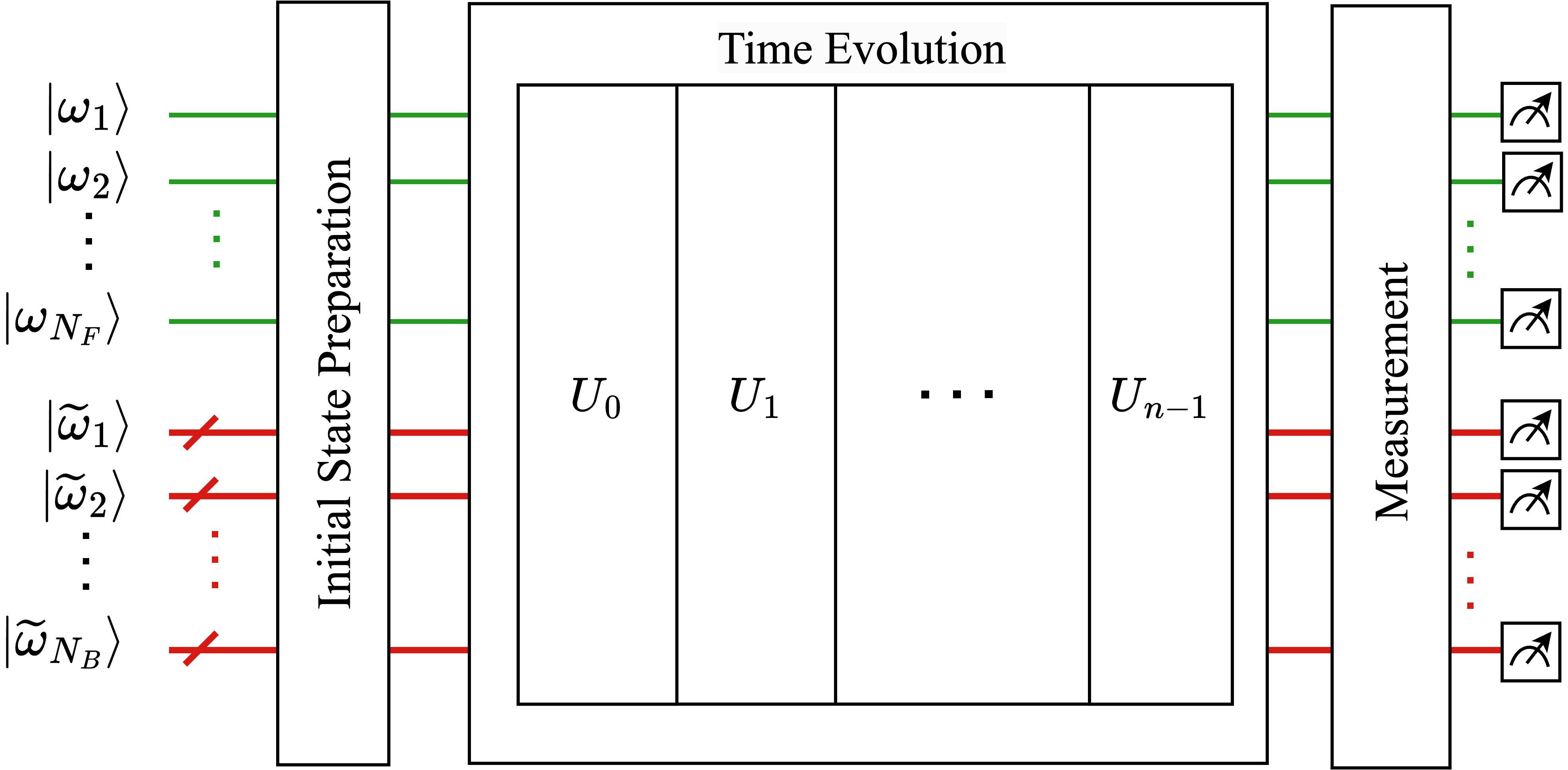}
    }
    \caption{Schematic illustrations of scalable quantum simulation for jet probe scattering with the medium. \textbf{a)} Spacetime diagram of the hard probe and background field. \textbf{b)} Unitary time evolution involving quantum inference. \textbf{c)} Qubit mapping scheme that uses qubits/registers to encode mode occupancies. \textbf{d)} Quantum circuit for simulating jet evolution.
    }
\end{figure}

The theoretical formulation of a high-energy parton evolving through an external background field (representing the medium) can be well-described by the non-perturbative light-front (LF) Hamiltonian formalism, which closely follows the established Discretized Light-Cone Quantization~\cite{Brodsky:1997de} and Basis Light-Front Quantization approaches~\cite{1stBLFQ} in the LF community.
In the spacetime diagram, we consider a high-energy jet (either quark or gluon) moving in the $x^+=t+z$ direction that scatters off a background field (either cold nuclear matter from a high-energy nucleus or the dense medium created in HIC), moving in the $x^-=t-z$ direction, as illustrated in Fig.~\ref{fig:spacetime}. In this setup, jet is treated as a fully quantum state governed by QCD dynamics with the nucleus/medium described by a background field for real-time evolution. The Lagrangian for the process under consideration is the QCD Lagrangian with an external background gluon field $\mathcal{A}^\mu$ in addition to the quantum gauge field $ A^\mu$. The subsequently derived LF Hamiltonian $P^-$ consists of the standard QCD Hamiltonian $P^-_\textrm{QCD}$ in vacuum and the time($x^+$)-dependent medium interaction $V_{\mathcal{A}}(x^+)$ with the background field $\mathcal{A}$~\cite{Li:2021zaw}.
% $\mathcal{A}$ adpoted from the McLerran–Venugopalan (MV) model~\cite{McLerran:1993ni}.
% Written out explicity, $P^-_{\text{KE}}$ is the kinetic energy term, $V_{qg}$, $V_{ggg}$, and $V_{gggg}$ are the quark-gluon and multi-gluon interaction vertices, $W_g, W_f$ are the gluon and fermion instantaneous terms, and $V_{q\mathcal{A}}(x^+)$ and $V_{g\mathcal{A}}(x^+)$ are the quark-field and gluon-field interactions with the background .
% The LF Hamiltonian $P^-$ is derived through the Legendre transformation~\cite{Brodsky:1997de, Li:2021zaw},
% \begin{equation}
%     \label{eq:Hamiltonian}
%     P^-(x^+) =P^-_\textrm{QCD}+ V_{\mathcal{A}}(x^+) \equiv P^-_{\text{KE}} + V_{qg}+V_{ggg}+V_{gggg}+W_{g} + W_{f} + V_{q\mathcal{A}}(x^+) + V_{g\mathcal{A}}(x^+)\;,
% \end{equation}
% which consists of the standard QCD Hamiltonian $P^-_\textrm{QCD}$ in vacuum and the interactions $V_{\mathcal{A}}(x^+)$ with the background field $\mathcal A$. Written out explicity, $P^-_{\text{KE}}$ is the kinetic energy term, $V_{qg}$, $V_{ggg}$, and $V_{gggg}$ are the quark-gluon and multi-gluon interaction vertices, $W_g, W_f$ are the gluon and fermion instantaneous terms, and $V_{q\mathcal{A}}(x^+)$ and $V_{g\mathcal{A}}(x^+)$ are the quark-field and gluon-field interactions with the background field $\mathcal{A}$ adpoted from the McLerran–Venugopalan (MV) model~\cite{McLerran:1993ni}.

The jet is formed originally from a highly energetic quark/gluon through a branching process dominated by gluon emission and absorption, before hadronization. Our focus is on simulating and studying the interplay between the branching process and medium interactions where multiple gluons are produced. 
% , we consider the Fock space expansion of the quark jet together with its relevant contributing terms of the Hamiltonian as follows,
% \begin{align}
% \label{eq:jet_states}
% \ket{q}_{\mathrm{jet}} =&\; \psi_q\ket{q}+\psi_{qg}\ket{qg}+\psi_{qgg}\ket{qgg}+\cdots\;,\\
% \label{eq:H_qjet}
% P_{q ~\mathrm{jet}}^-(x^+)=&\;
%     K_q + K_g
%     + V_{qg}
%     + V_{\mathcal{A}}(x^+)
%     +V_{ggg}
%     +V_{gggg}
%     +W_{q}
%     +W_{g}\;.
% \end{align}
% Similarly for the gluon jets and other processes. See Ref.~\cite{Qian:2024gph} for detailed second quantized operators of these Hamiltonian terms. 
Dynamical evolution of the jet as a quantum many-body state is governed by the time evolution equation on the LF~\cite{Li:2020uhl}, 
% \begin{align}
%   \label{eq:ShrodingerEq}
%   i\frac{\partial}{\partial x^+}\ket{\psi;x^+}=\frac{1}{2}P^-(x^+)\ket{\psi;x^+}\;.
% \end{align}
whose solution at any later LF time $x^+$ is obtained via the time-ordered path integral, and subsequently approximated by a sequence of unitary evolution operators 
% \begin{align}\label{eq:ShrodingerEqSol}
% \ket{\psi;x^+}=&\mathcal{T}_+\exp\left[-\frac{i}{2}\int_0^{x^+}\text{d} z^+P^-(z^+)\right]\ket{\psi;0}\;, \\
% =&\lim_{n\to\infty}\prod^{n-1}_{k=0}\mathcal{T}_+ \exp\left[-\frac{i}{2}\int_{x_{k}^+}^{x_{k+1}^+}\diff z^+P^-(z^+)\right]\ket{\psi;0} \;,\\
% =& \lim_{n\to\infty}\prod^{n-1}_{k=0}\exp\left[-\frac{i}{2} P^-(x_k^+) \delta x^+\right]\ket{\psi;0} = \lim_{n\to\infty}\prod^{n-1}_{k=0}U(x_k^+;x_{k+1}^+)\ket{\psi;0}\label{eq:unitary_evo}
% \end{align}
% with step size $\delta x^+ \equiv x^+/n$ for total number of steps $n$ and 
that essentially sums up infinitely many quantum interference effects (see Fig.~\ref{fig:interference}).
Within each short timestep, the Hamiltonian is considered time-independent and is described by the unitary evolution operator $U_k=U(x_k^+;x_{k+1}^+)=\exp\left[-\frac{i}{2} P^-(x_k^+) \delta x^+\right]$ for $x^+_k=k\delta x^+$. This procedure enables one to access the quantum state information and compute subsequent physical observables.

% At high energies, the jet has momentum $p^+\gg p_\perp > p^- $ so the probe is highly localized and one can simplify the field's spacetime dependence to $\mathcal{A}^\mu(x^+, \boldsymbol{x})$. 

% \begin{figure}
%   \centering
  
%   \includegraphics[width=0.38\textwidth]{spacetime.png}
%   \caption{Schematic of a jet probe scattering off a nucleus in a spacetime diagram.
%   }
%  \label{fig:spacetime}
% \end{figure}

\subsection{Quantum simulation strategy}

Benchmark classical calculations extrapolated to the continuum have been performed on supercomputers successfully up to leading two Fock sectors~\cite{Li:2020uhl,Li:2023jeh,Li:2025wzq}. 
As dimension of the resulting Hilbert space grows exponentially with the maximum number of particles in a Fock state, classical computation is computationally expensive even on supercomputer clusters, so it would be natural and ideal to simulate the Hamiltonian evolution on quantum computers.
In this section, we discuss various key aspects of simulating the evolution of a jet probe in medium on a digital quantum computer, including Hilbert space truncation, mapping of physical degrees of freedom onto qubits, and algorithms for state preparation and time evolution. 

\textbf{Finite Hilbert space input} for the jet many-body state is the first step of quantum simulation. The QCD Fock states constituting the Hilbert space of the jet use
$q_j$, $\overline{q}_j$, and $g_j$ to represent the quark, antiquark, and gluon single-particle states, with subscript $j$ denoting the index of different mode. Within single-particle state, each such mode is identified with five quantum numbers, $\{k^+, k^x, k^y, \lambda, c(a)\}$, where $(k^+,k^x, k^y)$ are LF momenta, $\lambda$ the LF helicity, and $c=1,2,..., N_c$ ($a=1,2,..., N_c^2-1$) the color index of quark/antiquark (gluon). Numerically, one can impose the maximal value in the longitudinal direction by the harmonic resolution $K$ and the transverse by $N_\perp$, without truncation to the Fock sector expansion. Upon introducing the cutoffs, an arbitrary QCD Fock state is discretized:
\begin{align}
\label{eq:fock_general_modes}
\ket{
q_1,\,q_2,\ldots;
\overline{q}_1,\,\overline{q}_2,\ldots;
g_1,\,g_2,\ldots
}
\rightarrow\ket{
q_1^{\occf_1},
\ldots,q_{\modesq}^{\occf_{\modesq}};
\overline{q}_1^{\overline{\occf}_1},
\ldots,\overline{q}_{\modesa}^{\overline{\occf}_{\modesa}};
g_1^{\occb_1},
\ldots,g_{\modesg}^{\occb_{\modesg}}
}\;,
\end{align}
with he total number of quark modes with distinct quantum numbers denoted by 
 $\modesq$, and analogously $\modesa$ for antiquarks and $\modesg$ for gluons. These numbers scale polynomially to the product of momentum space cutoffs, given by $K (2 N_\perp)^2$.
The variables $\occf_j, \overline{\occf}_j=0,1$ and $\occb_j=0,1,2,\ldots,\occmax_j$ represent the occupancies of fermionic and bosonic modes, respectively.

\textbf{Qubit encoding} refers to a mapping from the physical states to those of a multi-qubit system (spin chains). Several encoding strategies are available and have been explored, such as the basis state encoding~\cite{Barata:2022wim,Barata:2023clv,Wu:2024adk,Yao:2022eqm}, direct encoding~\cite{Qian:2024gph}, and compact encoding~\cite{Kreshchuk:2020dla}. In the case of direct encoding scheme, quantum registers are enumerated in the same way as the single-particle states to store the occupancies of corresponding modes in a Fock state; see Fig.~\ref{fig:encoding}. Importantly, compared to earlier basis state encoding, this direct encoding, based on second-quantized Hamiltonian, is efficient, scalable, and free from classical preparation of Pauli strings. The fermionic and bosonic creation/annihilation operators in the Hamiltonians are directly mapped into quantum gates on the circuit, with Jordan-Wigner or Bravyi-Kitaev encoding for fermionic operators ($b_\beta^\dagger,b_\beta$) on $N_F = J_q + J_{\bar{q}}$ qubits and binary/Gray/unary for bosonic ($a_\beta^\dagger,a_\beta$) on $N_B = J_g$ quantum registers.
% of various sizes.
% ranging from $\lceil\log_2(\occmax_j+1)\rceil$ in binary or Gray encodings to $(\occmax_j + 1)$ in unary.

\textbf{Simulation strategy} is the key to quantum simulation. Jet state preparation in this setup can be either a zero momentum state, which is useful to observe jet momentum broadening, or an eigenstate of the free $P^-_\textrm{QCD}$ Hamiltonian, which may be treated as a dressed particle traveling from $x^+=-\infty$. Various time evolution techniques can be used to evolve such states and are generally divided into Trotter methods and post-Trotter methods. Trotter methods use product formula that approximates exponential of the full Hamiltonian into a product of exponents of individual terms for each time step $t$, where its total gate cost scales as $\mathcal{O}(t^{1+\frac{1}{p}}\epsilon^{-\frac{1}{p}})$ at $p$-th order with precision $\epsilon$. Post-Trotter algorithms such as truncated Taylor series~\cite{Wu:2024adk} and quantum signal processing, may be considered, which typically involves constructing Block Encoding subcircuits, more suitable for improved precision over large-scale simulation on fault-tolerant quantum computers. 

\textbf{Measurement protocol} is the last stage of quantum simulation where information about the specific observable is extracted from the final state of evolution. It can be achieved by various approaches such as the quantum phase estimation, classical shadow, and VQE-type measurements. Jet observables of interests such as the transverse momentum broadening and gluon number can be expressed in terms of diagonal operators $a^\dagger_\beta a_\beta$ and $b^\dagger_\beta b_\beta$, whose expectation values are directly extracted from standard measurement. Owing to the stochastic nature of the classical background field (typically adopted from the McLerran–Venugopalan model), multiple field configurations are sampled to calculate expectation averages/deviation for the physical observables, very much analogous to conducting HEP experiments in a lab. A full schematic of the quantum circuit to simulate jets is provided in Fig.~\ref{fig:circuit}
.
\subsection{Recent progress}

In this section, we briefly review recent progress in simulating jets with quantum computing techniques. The first strategy to compute jet quenching parameter $\hat{q}$ was introduced in 2021 and subsequently demonstrated on classical quantum simulator along with experimental efforts on available NISQ hardware~\cite{Barata:2021yri, Barata:2022wim}. Using time-dependent Hamiltonian formalism, the simulation results are not only in good agreement with analytical analysis but also go beyond the eikonal limit to finite kinetic energy. Later on, more prototype studies have followed and extended the investigation to more than one Fock sector, such as $\ket{q}+\ket{qg}$ for the quark jet~\cite{Barata:2023clv}, and $\ket{g} + \ket{gg}$ for the gluon jet~\cite{Yao:2022eqm}. For the quark jet, one is able to observe sizable sub-eikonal effects, medium-induced modification to the gluon emissions, and super-linear exponential growth of entropy, verified in part by classical studies. Lately, post-Trotter method utilizing truncated Taylor series was proposed to improve the study within the leading $\ket{q}$ sector~\cite{Wu:2024adk}. Finally, a unified, systematic treatment of the quantum strategy to study jet phenomenology using direct encoding scheme was proposed and pushed the simulation to three gluons~\cite{Qian:2024gph}. Note, at present stage, all these numerical efforts are still limited on a small lattice space and largely performed using classical simulators of various backend techniques ranging from statevector, shot simulation, noisy simulation, to matrix product state methods.

\begin{figure}
    \centering
    \subfigure[\label{fig:res_p2}]{\includegraphics[width=0.47\textwidth]{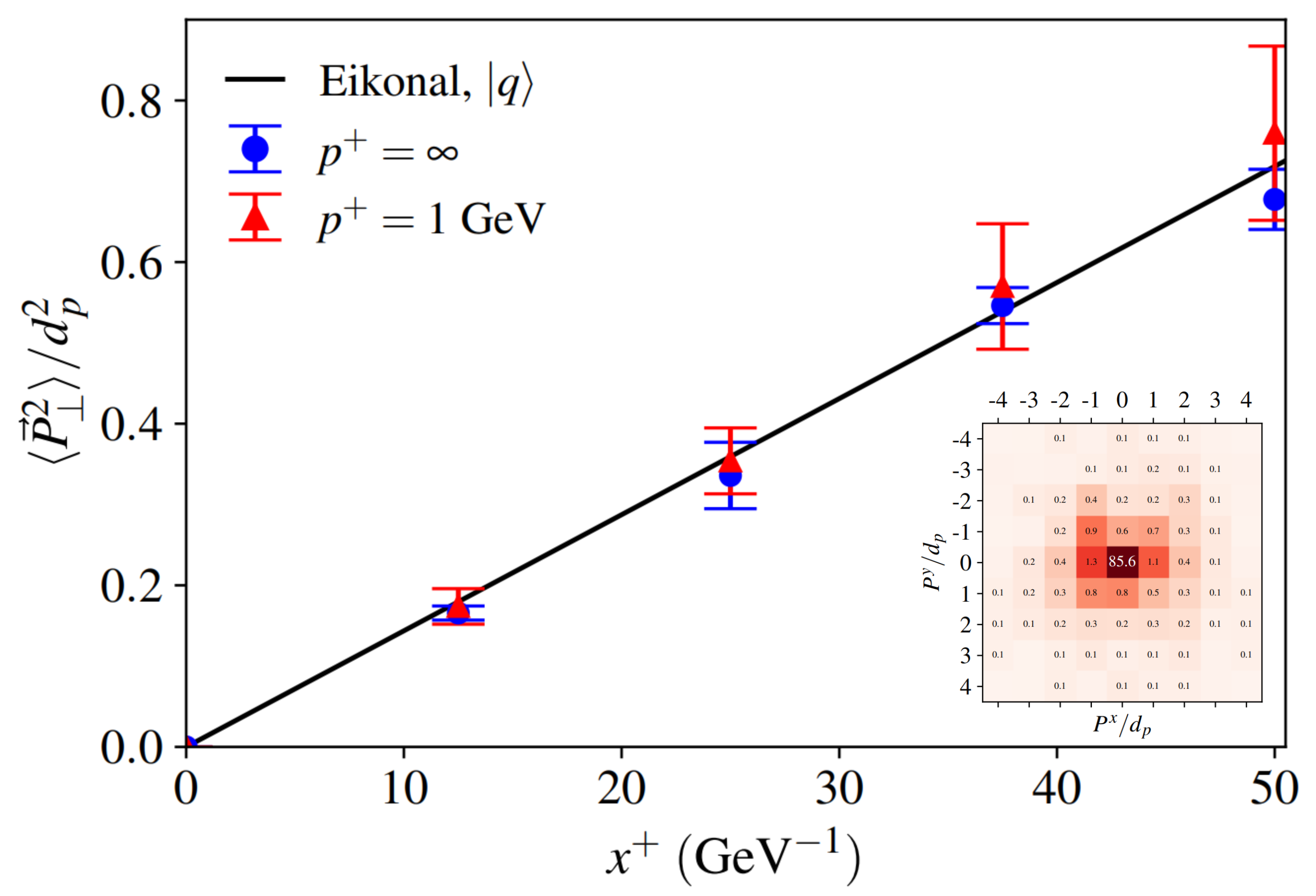}
    }\quad
    \subfigure[\label{fig:res_Ng}]{\includegraphics[width=0.47\textwidth]{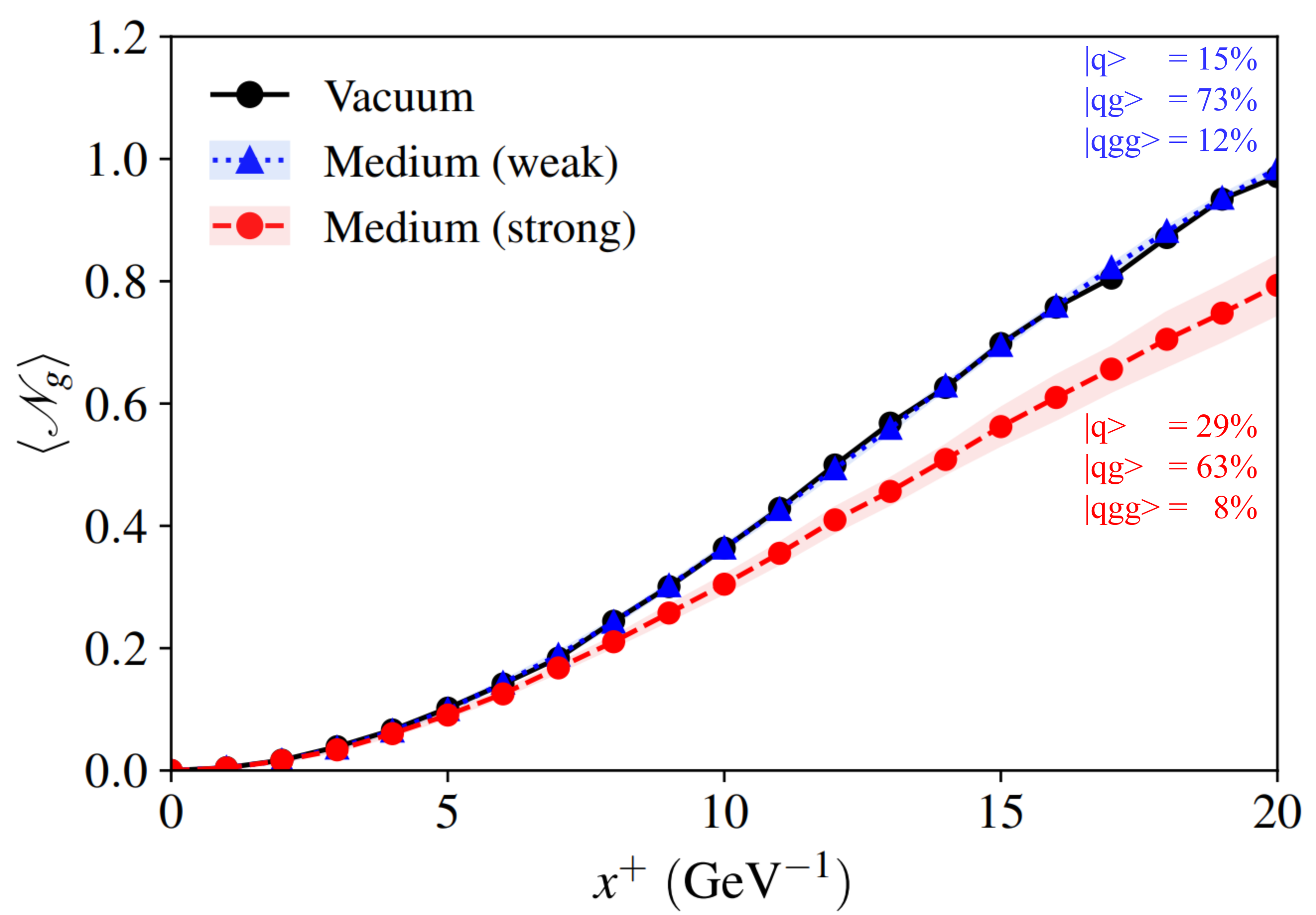}
    }
    \caption{Selected results of the quark jet in medium extracted from quantum circuits. (a) Transverse momentum broadening in the leading $|q\rangle$ Fock sector. The inset shows final probability distribution for the transverse momenta. (b) Medium modification to gluon number in $|q\rangle+|qgg\rangle+|qgg\rangle$.
    }
\end{figure}

As demonstration, we highlight some of the recent results of simulations for quark jet using direct encoding with the matrix product state simulator up to 128 qubits~\cite{Qian:2024gph}. In Fig.~\ref{fig:res_p2}, we study the  transverse momentum $\braket{\vec{P}^2_\perp; x^+}$ of a quark jet propagating through medium in the leading $\ket{q}$ Fock space. This medium-averaged expectation is related to the jet quenching parameters $\hat{q}$ via $\hat{q}L_\eta\equiv \braket{\vec{P}^2_\perp; x^+=L_\eta}-\braket{\vec{P}^2_\perp; 0}$ after total $L_\eta$ 
time evolution. Starting with a $\braket{\vec{P}^2_\perp; 0}=0$ initial quark state, the value of $\hat{q}$ extracted is found comparable to the analytical Wilson line result in the eikonal limit. This successfully illustrates single parton diffusion due to multiple medium interactions on the amplitude level. In Fig.~\ref{fig:res_Ng}, we study the medium modification to gluons radiation in the jet scattering, which is conveniently quantified by the gluon number operator $\mathcal{N}_g \equiv \sum_{\beta} a^\dagger_\beta a_\beta$. From the simulation, we see that a comparable medium could suppress gluon number produced in the probe at early stage due to quantum interference, which is also found in Refs.~\cite{Barata:2023clv, Yao:2022eqm}, though it remains an open question on a more realistic lattice.

% To minimize computational resource, we implement two simplifications in our simulations: we take $N_c=2$ for the color and fix the light-front helicity of each single-particle mode to the up configuration. 
% For the Trotterized simulations, we employ a finite timestep, which we have verified for numerical convergence by testing the results with decreasing timesteps. 

\section{Summary and future prospects}
In summary, QC presents a promising frontier for addressing some of the most challenging problems in high-energy physic, especially for HIC, from enhancing experimental data analysis with quantum algorithms to enabling detailed simulations of jet dynamics. See Refs.~\cite{Bauer:2022hpo, DiMeglio:2023nsa, Banuls:2019bmf} for an introduction and overview of many more works that are not mentioned.

Quantum computation approaches are thriving for studying jets in heavy ion collisions with application to enhance experimental measurement and several theoretical options available, such as lattice QFT approaches, open quantum system, and light-front Hamiltonian. Successful efforts are made to understand jets as a quantum many-body objects and its phenomenology under different contents. It remains important and interesting to extend current studies to describe more realistic scenarios with large scale simulation by considering medium anistropicity, initial stage effects, and particle multiplicity.

% Looking ahead, further research is needed to refine these simulation techniques and integrate them with real-world experimental setups. The potential for QC to revolutionize our understanding of heavy-ion collisions is immense, promising new discoveries and a deeper understanding of the fundamental forces that govern our universe.

% \wq{lastly, mention gluon jet, antenna, anistropic medium, initial stage, ... parton shower...}

\vspace{3mm}
\section*{Acknowledgments}
The author acknowledge useful discussions with João Barata, Michael Kreshchuk, Meijian Li, Carlos A. Salgado, James P. Vary, and Bin Wu for
their helpful and valuable discussions. This work is supported by the European Research Council under project ERC-2018-ADG-835105 YoctoLHC; by Maria de Maeztu excellence unit grant CEX2023-001318-M and project PID2020-119632GB-I00 funded by MICIU/AEI/10.13039/501100011033; by ERDF/EU; by the Marie Sklodowska-Curie Actions Postdoctoral Fellowships under Grant No. 101109293; by Xunta de Galicia (CIGUS Network of Research Centres).  

\bibliography{ref}

\begin{thebibliography}{37}

\bibitem{Bauer:2022hpo}
C.W. Bauer et~al., {Quantum Simulation for High-Energy Physics}, PRX Quantum \textbf{4}, 027001 (2023), \texttt{2204.03381}. \doiwoc{10.1103/PRXQuantum.4.027001}

\bibitem{DiMeglio:2023nsa}
A.~Di~Meglio et~al., {Quantum Computing for High-Energy Physics: State of the Art and Challenges}, PRX Quantum \textbf{5}, 037001 (2024), \texttt{2307.03236}. \doiwoc{10.1103/PRXQuantum.5.037001}

\bibitem{Banuls:2019bmf}
M.C. Ba\~nuls et~al., {Simulating Lattice Gauge Theories within Quantum Technologies}, Eur. Phys. J. D \textbf{74}, 165 (2020), \texttt{1911.00003}. \doiwoc{10.1140/epjd/e2020-100571-8}

\bibitem{Pires:2020urc}
D.~Pires, Y.~Omar, J.a. Seixas, {Adiabatic quantum algorithm for multijet clustering in high energy physics}, Phys. Lett. B \textbf{843}, 138000 (2023), \texttt{2012.14514}. \doiwoc{10.1016/j.physletb.2023.138000}

\bibitem{Delgado:2022snu}
A.~Delgado, J.~Thaler, {Quantum annealing for jet clustering with thrust}, Phys. Rev. D \textbf{106}, 094016 (2022), \texttt{2205.02814}. \doiwoc{10.1103/PhysRevD.106.094016}

\bibitem{Gianelle:2022unu}
A.~Gianelle, P.~Koppenburg, D.~Lucchesi, D.~Nicotra, E.~Rodrigues, L.~Sestini, J.~de~Vries, D.~Zuliani, {Quantum Machine Learning for b-jet charge identification}, JHEP \textbf{08}, 014 (2022), \texttt{2202.13943}. \doiwoc{10.1007/JHEP08(2022)014}

\bibitem{Lamm:2019uyc}
H.~Lamm, S.~Lawrence, Y.~Yamauchi (NuQS), {Parton physics on a quantum computer}, Phys. Rev. Res. \textbf{2}, 013272 (2020), \texttt{1908.10439}. \doiwoc{10.1103/PhysRevResearch.2.013272}

\bibitem{LiTianyin:2021kcs}
T.~Li, X.~Guo, W.K. Lai, X.~Liu, E.~Wang, H.~Xing, D.B. Zhang, S.L. Zhu (QuNu), {Partonic collinear structure by quantum computing}, Phys. Rev. D \textbf{105}, L111502 (2022), \texttt{2106.03865}. \doiwoc{10.1103/PhysRevD.105.L111502}

\bibitem{Banuls:2024oxa}
M.C. Ba\~nuls, K.~Cichy, C.J.D. Lin, M.~Schneider, {Parton Distribution Functions in the Schwinger Model with Tensor Networks}, in \emph{{41st International Symposium on Lattice Field Theory}} (2024), \texttt{2409.16996}

\bibitem{Kang:2025xpz}
Z.B. Kang, N.~Moran, P.~Nguyen, W.~Qian, {Partonic distribution functions and amplitudes using tensor network methods} (2025), \texttt{2501.09738}.

\bibitem{Grieninger:2024axp}
S.~Grieninger, I.~Zahed, {Quasifragmentation functions in the massive Schwinger model}, Phys. Rev. D \textbf{110}, 116009 (2024), \texttt{2406.01891}. \doiwoc{10.1103/PhysRevD.110.116009}

\bibitem{LiTianyin:2024nod}
T.~Li, H.~Xing, D.B. Zhang, {Simulating Parton Fragmentation on Quantum Computers} (2024), \texttt{2406.05683}.

\bibitem{Bauer:2021gup}
C.W. Bauer, M.~Freytsis, B.~Nachman, {Simulating Collider Physics on Quantum Computers Using Effective Field Theories}, Phys. Rev. Lett. \textbf{127}, 212001 (2021), \texttt{2102.05044}. \doiwoc{10.1103/PhysRevLett.127.212001}

\bibitem{Bepari:2021kwv}
K.~Bepari, S.~Malik, M.~Spannowsky, S.~Williams, {Quantum walk approach to simulating parton showers}, Phys. Rev. D \textbf{106}, 056002 (2022), \texttt{2109.13975}. \doiwoc{10.1103/PhysRevD.106.056002}

\bibitem{Hebenstreit:2013baa}
F.~Hebenstreit, J.~Berges, D.~Gelfand, {Real-time dynamics of string breaking}, Phys. Rev. Lett. \textbf{111}, 201601 (2013), \texttt{1307.4619}. \doiwoc{10.1103/PhysRevLett.111.201601}

\bibitem{Florio:2023dke}
A.~Florio, D.~Frenklakh, K.~Ikeda, D.~Kharzeev, V.~Korepin, S.~Shi, K.~Yu, {Real-Time Nonperturbative Dynamics of Jet Production in Schwinger Model: Quantum Entanglement and Vacuum Modification}, Phys. Rev. Lett. \textbf{131}, 021902 (2023), \texttt{2301.11991}. \doiwoc{10.1103/PhysRevLett.131.021902}

\bibitem{Farrell:2024fit}
R.C. Farrell, M.~Illa, A.N. Ciavarella, M.J. Savage, {Quantum simulations of hadron dynamics in the Schwinger model using 112 qubits}, Phys. Rev. D \textbf{109}, 114510 (2024), \texttt{2401.08044}. \doiwoc{10.1103/PhysRevD.109.114510}

\bibitem{Casalderrey-Solana:2007knd}
J.~Casalderrey-Solana, C.A. Salgado, {Introductory lectures on jet quenching in heavy ion collisions}, Acta Phys. Polon. B \textbf{38}, 3731 (2007), \texttt{0712.3443}.

\bibitem{Barata:2025hgx}
J.a. Barata, E.~Rico, {Real-time simulation of jet energy loss and entropy production in high-energy scattering with matter} (2025), \texttt{2502.17558}.

\bibitem{Barata:2024apg}
J.a. Barata, S.~Mukherjee, {Probing Celestial Energy and Charge Correlations through Real-Time Quantum Simulations: Insights from the Schwinger Model} (2024), \texttt{2409.13816}.

\bibitem{Lee:2024jnt}
K.~Lee, F.~Turro, X.~Yao, {Quantum computing for energy correlators}, Phys. Rev. D \textbf{111}, 054514 (2025), \texttt{2409.13830}. \doiwoc{10.1103/PhysRevD.111.054514}

\bibitem{DeJong:2020riy}
W.A. De~Jong, M.~Metcalf, J.~Mulligan, M.~P\l{}osko\'n, F.~Ringer, X.~Yao, {Quantum simulation of open quantum systems in heavy-ion collisions}, Phys. Rev. D \textbf{104}, 051501 (2021), \texttt{2010.03571}. \doiwoc{10.1103/PhysRevD.104.L051501}

\bibitem{deJong:2021wsd}
W.A. de~Jong, K.~Lee, J.~Mulligan, M.~P\l{}osko\'n, F.~Ringer, X.~Yao, {Quantum simulation of nonequilibrium dynamics and thermalization in the Schwinger model}, Phys. Rev. D \textbf{106}, 054508 (2022), \texttt{2106.08394}. \doiwoc{10.1103/PhysRevD.106.054508}

\bibitem{Barata:2021yri}
J.a. Barata, C.A. Salgado, {A quantum strategy to compute the jet quenching parameter $\hat{q}$}, Eur. Phys. J. C \textbf{81}, 862 (2021), \texttt{2104.04661}. \doiwoc{10.1140/epjc/s10052-021-09674-9}

\bibitem{Barata:2022wim}
J.a. Barata, X.~Du, M.~Li, W.~Qian, C.A. Salgado, {Medium induced jet broadening in a quantum computer}, Phys. Rev. D \textbf{106}, 074013 (2022), \texttt{2208.06750}. \doiwoc{10.1103/PhysRevD.106.074013}

\bibitem{Barata:2023clv}
J.a. Barata, X.~Du, M.~Li, W.~Qian, C.A. Salgado, {Quantum simulation of in-medium QCD jets: Momentum broadening, gluon production, and entropy growth}, Phys. Rev. D \textbf{108}, 056023 (2023), \texttt{2307.01792}. \doiwoc{10.1103/PhysRevD.108.056023}

\bibitem{Qian:2024gph}
W.~Qian, M.~Li, C.A. Salgado, M.~Kreshchuk, {Efficient Quantum Simulation of QCD Jets on the Light Front} (2024), \texttt{2411.09762}.

\bibitem{Wu:2024adk}
S.~Wu, W.~Du, X.~Zhao, J.P. Vary, {Efficient and precise quantum simulation of ultrarelativistic quark-nucleus scattering}, Phys. Rev. D \textbf{110}, 056044 (2024), \texttt{2404.00819}. \doiwoc{10.1103/PhysRevD.110.056044}

\bibitem{Yao:2022eqm}
X.~Yao, {Quantum Simulation of Light-Front QCD for Jet Quenching in Nuclear Environments} (2022), \texttt{2205.07902}.

\bibitem{Castro:2025ocx}
N.F. Castro, J.G. Milhano, M.G.J.a. Oliveira, {Jet evolution in a quantum computer: quark and gluon dynamics} (2025), \texttt{2502.03431}.

\bibitem{Brodsky:1997de}
S.J. Brodsky, H.C. Pauli, S.S. Pinsky, {Quantum chromodynamics and other field theories on the light cone}, Phys. Rept. \textbf{301}, 299 (1998), \texttt{hep-ph/9705477}. \doiwoc{10.1016/S0370-1573(97)00089-6}

\bibitem{1stBLFQ}
J.P. Vary, H.~Honkanen, J.~Li, P.~Maris, S.J. Brodsky, A.~Harindranath, G.F. de~Teramond, P.~Sternberg, E.G. Ng, C.~Yang, {Hamiltonian light-front field theory in a basis function approach}, Phys. Rev. \textbf{C81}, 035205 (2010), \texttt{0905.1411}. \doiwoc{10.1103/PhysRevC.81.035205}

\bibitem{Li:2021zaw}
M.~Li, T.~Lappi, X.~Zhao, {Scattering and gluon emission in a color field: A light-front Hamiltonian approach}, Phys. Rev. D \textbf{104}, 056014 (2021), \texttt{2107.02225}. \doiwoc{10.1103/PhysRevD.104.056014}

\bibitem{Li:2020uhl}
M.~Li, X.~Zhao, P.~Maris, G.~Chen, Y.~Li, K.~Tuchin, J.P. Vary, {Ultrarelativistic quark-nucleus scattering in a light-front Hamiltonian approach}, Phys. Rev. D \textbf{101}, 076016 (2020), \texttt{2002.09757}. \doiwoc{10.1103/PhysRevD.101.076016}

\bibitem{Li:2023jeh}
M.~Li, T.~Lappi, X.~Zhao, C.A. Salgado, {Momentum broadening of an in-medium jet evolution using a light-front Hamiltonian approach}, Phys. Rev. D \textbf{108}, 036016 (2023), \texttt{2305.12490}. \doiwoc{10.1103/PhysRevD.108.036016}

\bibitem{Li:2025wzq}
M.~Li, T.~Lappi, X.~Zhao, C.A. Salgado, {Scattering and gluon emission of physical quarks in a SU(3) colored field} (2025), \texttt{2504.07162}.

\bibitem{Kreshchuk:2020dla}
M.~Kreshchuk, W.M. Kirby, G.~Goldstein, H.~Beauchemin, P.J. Love, {Quantum simulation of quantum field theory in the light-front formulation}, Phys. Rev. A \textbf{105}, 032418 (2022), \texttt{2002.04016}. \doiwoc{10.1103/PhysRevA.105.032418}

\end{thebibliography}

\end{document}